%% file: ms.tex
\documentclass[12pt,manuscript,doublespace]{aastex}

\usepackage{natbib}                
\usepackage{verbatim}
\usepackage{subfigure}
\usepackage{epsfig}

\def\arcsec{\ifmmode '' \else $''$\fi}

\def\arcsecpoint{\ifmmode ''\!. \else $''\!.$\fi}

\def\kms{\ifmmode {\rm km\ s}^{-1} \else km s$^{-1}$\fi}
\def\Msun{\ifmmode {\rm M}_{\odot} \else M$_{\odot}$\fi}
\def\Lsun{\ifmmode {\rm L}_{\odot} \else L$_{\odot}$\fi}
\def\Zsun{\ifmmode {\rm Z}_{\odot} \else Z$_{\odot}$\fi}

\def\ergscm2{ergs\,s$^{-1}$\,cm$^{-2}$}
\def\icm3{{\rm cm}^{-3}}
\def\icm2{{\rm cm}^{-2}}
\def\qo{\ifmmode q_{\rm o} \else $q_{\rm o}$\fi}
\def\Ho{\ifmmode H_{\rm o} \else $H_{\rm o}$\fi}
\def\ho{\ifmmode h_{\rm o} \else $h_{\rm o}$\fi}

\def\vFWHM{\ifmmode v_{\mbox{\tiny FWHM}} \else
            $v_{\mbox{\tiny FWHM}}$\fi}
\def\CCF{\ifmmode F_{\it CCF} \else $F_{\it CCF}$\fi}
\def\ACF{\ifmmode F_{\it ACF} \else $F_{\it ACF}$\fi}
\def\Halpha{\ifmmode {\rm H}\alpha \else H$\alpha$\fi}
\def\Hbeta{\ifmmode {\rm H}\beta \else H$\beta$\fi}
\def\Hgamma{\ifmmode {\rm H}\gamma \else H$\gamma$\fi}
\def\Hdelta{\ifmmode {\rm H}\delta \else H$\delta$\fi}
\def\Lya{\ifmmode {\rm Ly}\alpha \else Ly$\alpha$\fi}
\def\Lyb{\ifmmode {\rm Ly}\beta \else Ly$\beta$\fi}
\def\Lyg{\ifmmode {\rm Ly}\beta \else Ly$\gamma$\fi}

\def\feii{Fe\,{\sc ii}}

\def\ciii{\ifmmode {\rm C}\,{\sc iii} \else C\,{\sc iii}\fi}
\def\civ{\ifmmode {\rm C}\,{\sc iv} \else C\,{\sc iv}\fi}
\def\cv{\ifmmode {\rm C}\,{\sc v} \else C\,{\sc v}\fi}
\def\cvi{\ifmmode {\rm C}\,{\sc vi} \else C\,{\sc vi}\fi}

\def\oii{O\,{\sc ii}}
\def\oiii{O\,{\sc iii}}
\def\o5007{[O\,{\sc iii}]\,$\lambda5007$}

\def\mgi{Mg\,{\sc i}}
\def\mnii{Mn\,{\sc ii}}

\def\mgii{Mg\,{\sc ii}}

\def\siII{Si\,{\sc ii}}

\def\niII{Ni\,{\sc ii}}

\def\siv{S\,{\sc iv}}

\def\svi{S\,{\sc vi}}

\def\feii{Fe\,{\sc ii}}

\def\o{\o}

\def\gtorder{\mathrel{\raise.3ex\hbox{$>$}\mkern-14mu
             \lower0.6ex\hbox{$\sim$}}}
\def\ltorder{\mathrel{\raise.3ex\hbox{$<$}\mkern-14mu
             \lower0.6ex\hbox{$\sim$}}}
\def\proptwid{\mathrel{\raise.3ex\hbox{$\propto$}\mkern-14mu
             \lower0.6ex\hbox{$\sim$}}}

\begin{document}

\shortauthors{Dunn, et al.}
\shorttitle{Dust in FeLoBALs}

\title{Determining the Locations of Dust Sources in FeLoBAL Quasars}

\author{Jay P. Dunn\altaffilmark{1}, Branden Wasik\altaffilmark{1}, Christin L. Holtzclaw\altaffilmark{1}, David Yenerall\altaffilmark{1}, Manuel Bautista\altaffilmark{2}, Nahum Arav\altaffilmark{3}, Daniel Hayes\altaffilmark{1}, Max Moe\altaffilmark{4}, Luis C. Ho\altaffilmark{5,6}, S. Harper Dutton\altaffilmark{1}}

\altaffiltext{1}{Department of Physical Sciences, Georgia Perimeter College, Dunwoody, GA 30338, USA: jdunn@gpc.edu}
\altaffiltext{2}{Department of Physics, Western Michigan University, Kalamazoo, MI 49008-5252, USA}
\altaffiltext{3}{Department of Physics, Virginia Tech, Blacksburg, VA 24060, USA}
\altaffiltext{4}{Harvard-Smithsonian Center for Astrophysics, 60 Garden Street, MS-10, Cambridge, MA, 02138, USA}
\altaffiltext{5}{Kavli Institute for Astronomy and Astrophysics,
Peking University, Beijing 100871, China}
\altaffiltext{6}{The Observatories of the Carnegie Institution for Science,
813 Santa Barbara Street, Pasadena, CA 91101, USA}

\begin{abstract}

We conduct a spectroscopic search of quasars observed by the Sloan 
Digital Sky Survey (SDSS) with broad absorption line (BAL) troughs due to 
\mgii\ and troughs due to \feii\ that simultaneously
exhibit strong Balmer narrow emission lines. We find that in a redshift 
range of 0.4 $\le$ z $\le$ 0.9 approximately 23 of the 70 \mgii\ BALs and 
4 of a subset of 15 \feii\ BALs exhibit strong Balmer emission. We also
find significant fractions of \mgii\ BALs (approx. 23\%) and those 
\mgii\ BALs with \feii\ troughs (approx. 27\%) have strong continuum 
reddening, E(B$-$V) $\ge$ 0.1. From
measurements of the Balmer decrement in three objects, we find similarly 
significant reddening of the narrow emission line region in 3 of the 4 
objects; the narrow emission lines in the fourth object are 
not measurable. We also include one object in this study not taken from 
the SDSS sample that 
shows \feii\ absorption and strong narrow emission, but due to measurement 
uncertainty and low continuum reddening the comparison is consistent but 
inconclusive. We find a trend in both the \mgii\ and \feii\ BAL 
samples between the narrow emission line reddening and continuum reddening. 
Because the narrow line reddening is consistent with the continuum reddening 
in every object in the two SDSS samples, it suggests that the reddening 
sources in these objects likely exist at larger radial distances than the 
narrow line regions from the central nucleus.

\end{abstract}

\keywords{quasars: absorption lines, galaxies: evolution}

\input{Body.tex}


\bibliographystyle{apj}

\bibliography{ms}

\end{document}

%% file: Body.tex
\section{Introduction}

Mass outflows in quasars, seen as blueshifted broad absorption lines 
(BALs), have recently become a plausible explanation for quasar feedback 
\citep[e.g.,][]{2012MNRAS.420.1347F}. Feedback has been invoked to explain 
several phenomena such as the M$-\sigma$ relationship for supermassive black
holes (SMBH) and their host galaxies \citep[e.g.,][]{1998A&A...331L...1S,
2005Natur.433..604D} as well as the quenching of star formation in the host 
galaxy \citep[e.g.,][]{2013MNRAS.432.2639H,2013MNRAS.433.3297D}. The 
recent advancement is due in large part to studies 
of a subclass of BAL quasars that exhibit absorption troughs from both 
resonance and metastable state lines of low ionization species, 
namely from \feii\ and \feii*  \citep[i.e., FeLoBALs;][]{2008ApJ...688..108K,
2009ApJ...706..525M,2010ApJ...709..611D,2010ApJ...713...25B}
and in one case from high ionization lines 
\citep[\siv\ and \svi*;][]{2013ApJ...762...49B}.
The combination of these lines with photoionization modeling yields the 
density and therefore the distance and energy output of the outflows. 
These observational results have led to recent work to explain the physical 
mechanisms behind the driving of the gas 
\citep[e.g.,][]{2010ApJ...722..642O,2010MNRAS.406L..55D,2010ApJ...725..556S}. 

Unlike the more common high ionization BALs, which demonstrate strong 
absorption due to \civ\ and appear in approximately 20\% of all quasars 
\citep{2008MNRAS.386.1426K,2003AJ....125.1784H},
low ionization absorption troughs due to species like \mgii\ (LoBALs) and
\feii\ (FeLoBALs) are relatively uncommon in optical spectral surveys 
\citep[e.g.,][]{2000ApJS..126..133W,2002ApJS..141..267H,2010ApJ...725..556S}. 
Furthermore, many LoBALs \citep{2010ApJ...714..367Z} and FeLoBALs 
\citep{1991ApJ...373...23W,1992ApJ...390...39S,2003AJ....126.2594R,
2008ApJ...674...80U} display signs of large color excess due to dust 
extinction. The extinction commonly observed in these objects  
appears to be best described by a Small Magellanic Cloud (SMC) extinction 
curve \citep{2003AJ....126.1131R,2004AJ....128.1112H}. The combination of 
rarity and reddening properties of these 
objects has led studies to postulate about their relevance in quasar 
evolution. One plausible scenario is that FeLoBALs are normal BAL quasars 
with a specialized sight-line that grazes the edge of the putative dusty 
torus (see Hall et al. 2002). This simultaneously explains the strong 
reddening present in many objects and provides a source of opacity that 
decreases the ionizing flux impinging on the outflowing gas allowing for 
clouds to form singly ionized species (see Dunn et al. 2010 and references 
therein). An alternative scenario suggests that FeLoBALs represent an 
evolutionary state during a quasar’s lifetime \citep{
1996AJ....112...73E,1997ApJ...479L..93B,2008ApJS..175..356H,
2008ApJ...674...80U,2012ApJ...757...51G} where the quasar sheds its 
obscuring dust and gas. The dust source during this process could 
provide reddening at any distance from the nucleus.

While recent studies of FeLoBALs have determined distances 
for these outflows, the radial location of the dust source 
is unknown. This was especially relevant in the object QSO0318$-$0600, 
where we accurately determined the density of the 
outflowing gas due to the presence of troughs from metastable, 
excited-state lines (\feii, \siII, and \niII). Combining the density 
with photoionization modeling yielded a radial distance of several 
kiloparsecs from the nucleus for the outflowing material. Other studies 
have shown that large distances such as this appear in several FeLoBAL 
objects (e.g., SDSS~J0838+2955; Moe et al. 2009). 
Like many FeLoBALs, QSO0318$-$0600 is an extremely reddened quasar 
(see discussion in Dunn et al. 2010), where the impact of the dust on 
the full spectral energy distribution (SED) is significant. Thus, 
should the dust source be located between the outflow and the nucleus 
(similar to the torus scenario) then the ionizing flux reaching the 
outflowing gas would be severely diminished. This uncertainty in dust
location affects the distance determination and in turn the determined 
energy output by up to a factor of 4. 

We take the first steps to ascertain the radial location 
of dust in a sample of FeLoBALs, which has implications for the 
generation of low ionization species. We begin by discussing the 
method for spectral analysis in \S2 and determine the respective sample 
for this method in \S3. In \S4, we determine and compare the continuum 
and emission line reddenings for the objects, which provides the radial 
locations of the dust sources. Finally, we summarize the findings and 
discuss the implications in \S5. 

\section{Diagnostic Tool to Determine Radial Location}

As mentioned earlier, the sources of dust in LoBAL and FeLoBAL quasars 
could exist in two different radial distance regimes, either
interior or exterior to the outflowing gas with respect to the nucleus. 
To explain the commonly observed reddening in FeLoBALs, we assert 
logically plausible sources for the dust. For example, given the unified 
model of active galactic nuclei \citep[AGN,][]{1985ApJ...297..621A,
1995PASP..107..803U}, a grazing sightline past the edge of the torus in 
the vicinity of the AGN provides the observed reddening, which we 
refer to as scenario 1. The dust torus would provide the 
simplest simultaneous explanation for the reddening and the necessary 
decrease in ionizing flux required to generate low ionization species 
in these types of outflows. Other structures do exist such as nuclear 
dust spirals observed in spatially resolved low-z quasars and Seyfert 
galaxies that could potentially provide a reddening near the nucleus. 
In scenario 2, a source not necessarily associated with the AGN such 
as a galactic dust lane at kpc scale distances is the cause of the 
reddening. Without spatially resolved imaging of 
the host galaxy we cannot distinguish which specific structure is 
present, but we are able to determine 
whether the dust source lies close to the nucleus or at galactic 
distances.

Observationally, the torus in scenario 1 would redden 
both the continuum source and the broad emission lines (BELs) that arise 
from the broad line region (BLR) close to the accretion disk-SMBH system. 
By comparison, the narrow emission lines (NELs), which are generated in the 
narrow line region (NLR) at a much larger distance 
\citep[100s of parsecs in typical quasars;][]{1997iagn.book.....P}, 
would remain unextincted much like the work on Seyfert 1.8 and 1.9 galaxies 
by \citep{2010ApJ...725.1749T} and on the Seyfert 1 galaxy MCG$-6-30-15$ 
by \citep{1997MNRAS.291..403R}). In the second scenario, a large structure 
at a galactic scale would redden the continuum, BLR, and NLR by 
approximately similar amounts. Furthermore, a structure
at kpc scale distances would be statistically unlikely to cover only
one of the two AGN components. Thus, direct comparison of the continuum 
and NLR reddening demonstrates whether the reddening occurs interior 
(scenario 1) or exterior (scenario 2) to the NLR. A similarly 
reddened NLR compared to the continuum (i.e., E(B$-$V)$_{NLR} \approx$ 
E(B$-$V)$_{cont}$) suggests that the dust lies exterior to the NLR and 
cannot be explained by a dust torus. A reddening value of 
the continuum significantly larger than that of the NLR implies that 
the dust is in the vicinity of the AGN and is likely associated 
with the dust torus. 

It is important to note that the BELs 
should be as reddened as the continuum in either scenario, as the 
BLR lies interior to the torus. Ideally we would use the emission
ratios of the Balmer BELs in conjunction with the NELs to determine if
scenario 1 applies. However, it has been shown that 
BEL line ratios in AGN are not well described by case-B recombination 
in some objects \citep[e.g.][]{1981ApJ...250..478K,2004ApJ...606..749K}.
This primarily stems from high densities associated with the BLR region,
which leads to radiative transfer and collisional effects that impact
the description of recombination present in the gas. Because measurements
of the Balmer decrement of BELs are potentially poor approximations of
the extinction, we only compare the extinctions of the continuum source 
and the NEL.

Unlike the BELs, the NELs are adequately described by case-B recombination
as the number density is significantly lower ($10^4$ cm$^{-3} 
<$ n$_H < 10^6$ cm$^{-3}$), which implies that the case-B approximation is 
valid \citep{2006agna.book...O}. Furthermore, the ratio of H$\alpha$ to
H$\beta$ only spans the range 2.74$-$2.86 (due to density) at temperatures 
of approximately 10,000K, implying that the ratio is relatively insensitive
to temperature differences.

Assuming case-B, the reddening is defined as:
\begin{equation}
E(B-V)=\frac{2.21}{R_{H\alpha} - R_{H\beta}}~log\frac{2.76}{\frac{f_{H\alpha}}{f_{H\beta}}},
\end{equation}
where $R_{H\alpha}$ and $R_{H\beta}$ are the values of the SMC reddening 
curve at $H\alpha$ and $H\beta$, respectively, and $f_{H\alpha}$ and 
$f_{H\beta}$ are the peak fluxes of the emission lines. We use 2.76 as 
the intrinsic Balmer ratio as determined by case-B recombination (Osterbrock
\& Ferland 2006). While H$\alpha$ and H$\beta$ are the optimal lines for 
this process, in the majority of our sample we are only able to observe 
H$\beta$ and H$\gamma$ due to the combination of redshift and spectral 
range of the Sloan Digital Sky Survey (see \S3). Thus, we also use the 
ratio of H$\gamma$ to H$\beta$ (0.474, Osterbrock \& Ferland 2006), 
which likewise spans a relatively narrow range of 0.469$-$0.476 for 
case-B recombination.

\section{Survey}

\subsection{Sample Selection}

To employ the method outlined in Section 2, we require a sample of quasars 
that demonstrate FeLoBAL troughs as well as strong narrow line Balmer 
emission. To obtain the largest number of objects, we utilize the Sloan 
Digital Sky Survey (SDSS) spectroscopically observed quasars (Richards et. 
al 2009) through Data Release 7 \citep[DR7,][]{2010AJ...139.2360S}. Quasar
spectra in the SDSS-III catalog are unusable for this study as the spectra 
in that catalog are not accurately flux calibrated 
\citep{2013AJ....145...10D}. We access and download the calibrated spectra 
through the online Catalog Archive Server (CAS). 

We limit our search primarily by redshift. A redshift of z=0.45 is 
sufficient to redshift the \mgii\ ($\lambda\lambda$2796,2804) emission 
line into the spectral range 
of SDSS (approximately 3800 to 9200 \AA) and permit absorption trough 
detections for \mgii\ up to approximately 10,000 km s$^{-1}$. A redshift 
of 0.45 places the Balmer H$\alpha$ ($\lambda$6563) at 
$\approx$9500 \AA\ and out of the spectral range of SDSS, thus precluding 
simultaneous observation of BAL troughs and the H$\alpha$ emission line. 
We select an upper limit on the redshift determined by the 
Balmer H$\beta$ ($\lambda$4861) line, which redshifts out of the spectral 
range at approximately z=0.90. We further limit the sample by magnitude 
as objects fainter than r’ magnitude of 19 will have substantial noise 
that can potentially obscure weaker narrow Balmer emission lines. The 
resulting sample contains 8,511 quasars.

Next, we plot and visually inspect these spectra for blueshifted
\mgii\ and \feii\ BALs. To reduce the search time we examine
spectra for \feii\ troughs in those objects that contain \mgii\ BALs; 
previously these troughs have only been observed in objects that 
show \mgii\ absorption, which is typically easier to identify. 
We define BALs here to be any 
intrinsic absorption troughs in the quasar’s spectrum \citep[i.e., 
outflowing gas ejected by the AGN, see][]{1997AJ....113..136B}. 
Zhang et al. (2009) suggested that a velocity width of 1600 km s$^{-1}$ 
simultaneously maximizes the frequency of which BALs are intrinsic 
outflows and also maximizes completeness of a BAL sample. To this end, 
we only include objects with contiguous trough widths greater than 
2,000 km s$^{-1}$. A width of 2,000 km s$^{-1}$ is similar to the 
Absorption Index ($AI$) method used in Trump et al. (2006) and 
Hall et al.(2002). A width of 2,000 km s$^{-1}$ is also sufficient to 
blend the two members of the \mgii\ doublet and facilitate visual
inspection. This provides a conservative lower limit on objects with 
intrinsic absorption. Comparing to the Zhang et al. determinations, 
our sample is complete to slightly less than 75\%, but over 90\% of 
the objects are bona fide outflows tied to the quasar. Quasars in the 
lower redshift ranges only provide minimal velocity coverage of the 
UV~3 \feii\ multiplet ($\approx$$\lambda$2600). For the redshift range 
of 0.45$-$0.50, we detect only those BALs at low velocity (v$\sim$6000 
km s$^{-1}$ at z=0.50) and obtain only a lower limit on the number of 
objects with \feii\ BAL troughs.

The last criterion we impose on the sample is to select objects with \mgii\ 
and \feii\ BALs that show narrow emission lines due to H$\beta$ and/or 
H$\gamma$ ($\lambda$4341) in conjunction with [\oiii] $\lambda$5007 
emission. We characterize this as a distinct emission feature that has 
a maximum FWHM less than 1000 km s$^{-1}$. This value is consistent with 
the findings of \citealt{2003AJ....126.1131R} for [\oii] $\lambda$3727 
and [\oiii] emission lines in reddened quasars. The peak intensity 
must be distinguishable from the BEL (i.e., at least 3 $\sigma$ above 
the noise level at the BEL peak). 

\subsection{Survey Objects and Properties}

In our sample, we find 70 quasars (approximately 0.8\% of the full 8,511 
quasar sample) that show \mgii\ 
troughs matching the criteria established in the previous section. This 
is similar to other spectroscopically selected samples in the optical
(e.g., approximately 1.3\% via the $AI$ method in Trump et al. 2006). 
Of these 70 objects, 15 showed BALs attributed to \feii\ (0.2\%). To 
determine the presence of \feii\ troughs, we check for radial velocity 
agreement with \mgii\ as demonstrated in Figure 1 for a high resolution 
spectrum of SDSS~J0149$-$1016. Finally, only
4  of the 15 objects also demonstrate evidence of strong narrow emission 
due to hydrogen, [\oiii], and [\oii]. Figure 2 displays the SDSS spectra of 
these 4 objects (and 1 additional object explained below). The \mgii\ and 
\feii\ BAL troughs are evident in these
spectra. Furthermore, in all 4 cases, we find evidence of troughs due 
to excited, metastable state lines of \feii. We list the SDSS identifiers, 
r' magnitudes, and redshifts for all 70 of the BAL quasars in Table 1 and 
indicate which objects in the sample show troughs due to \feii\ and strong 
narrow H$\beta$ emission. Of the 4 FeLoBALs with detectable H$\beta$ 
emission, only SDSS~J~0802+5513 also has a detectable H$\gamma$ narrow 
emission line.

\begin{figure}[!h]
  \centering \includegraphics[angle=90,width=0.9\textwidth]
  {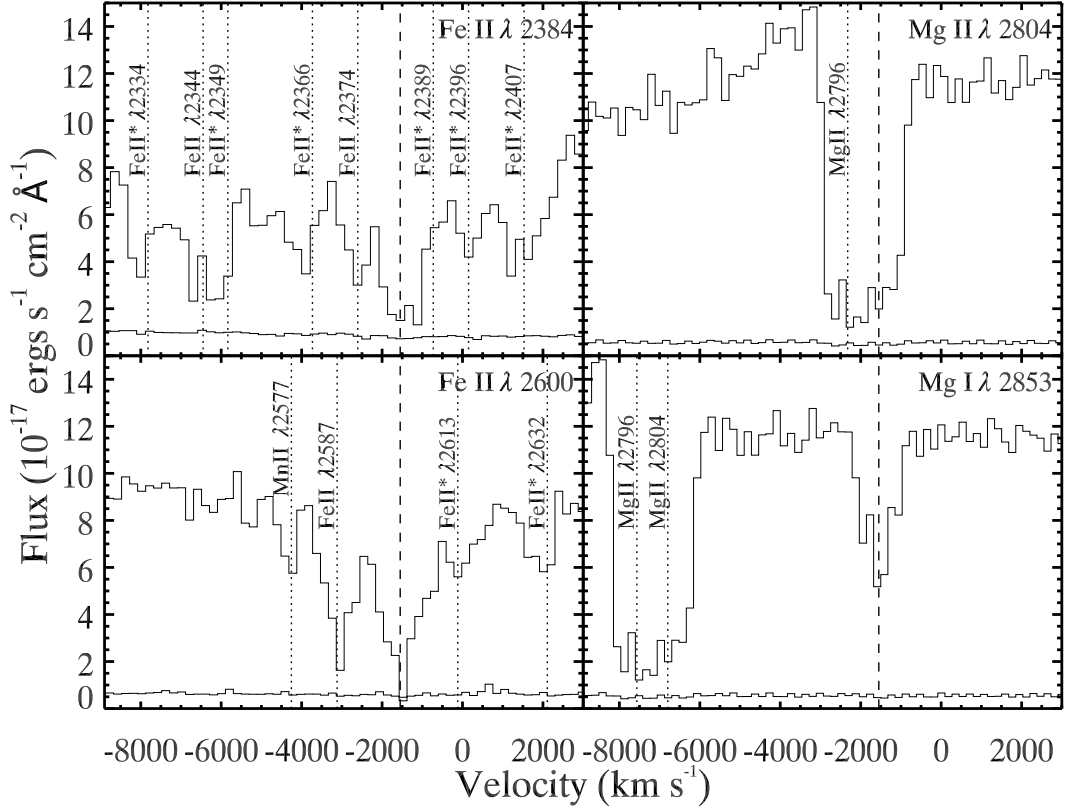}\\
  \caption[]
  {\footnotesize Absorption trough identifications for \mgi,
\mgii, \mnii\ and several resonance and excited state (noted
with an asterisk) \feii\ lines in the SDSS spectrum of
SDSS~J0802+5513. The primary absorption trough, marked
with a dashed line, occurs at approximately $-$1600 km s$^{-1}$.
Each panel is plotted in velocity from the respective ions
listed in the upper right. Other notable lines are indicated
with dotted lines at shifts of 1600 km s$^{-1}$ from their
respective rest wavelengths.}
  \label{f1}
\end{figure}

\clearpage
\begin{rotate}
\begin{figure}
\centering \includegraphics[angle=90,width=1.0\textwidth]{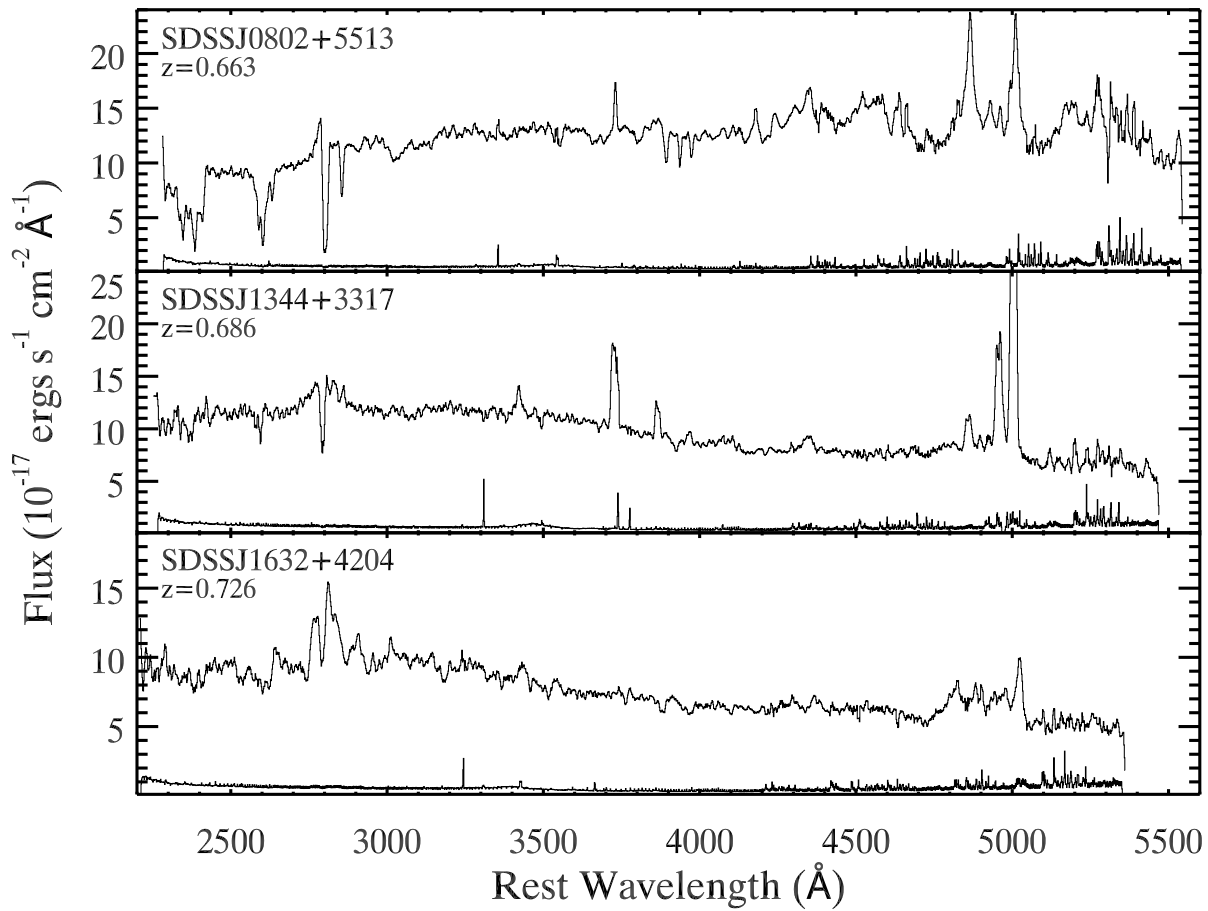}\\
\caption[]
{\footnotesize Spectra, plotted in the rest frame, 
for objects demonstrating BALs due to \mgii, with troughs due to 
\feii, and strong NELs ([\oiii] and H$\beta$). The object
SDSS names and SDSS measured redshifts are listed in the upper
left of each panel. The first 4 spectra were taken by SDSS, while
the final spectrum (SDSS~J0149$-$1016) was taken with the Magellan
MIKE spectrograph. The flux uncertainties are plotted beneath 
each spectrum with black histograms.}
  \label{f2}
\end{figure}
\end{rotate}

\clearpage
\begin{figure}
\plotone{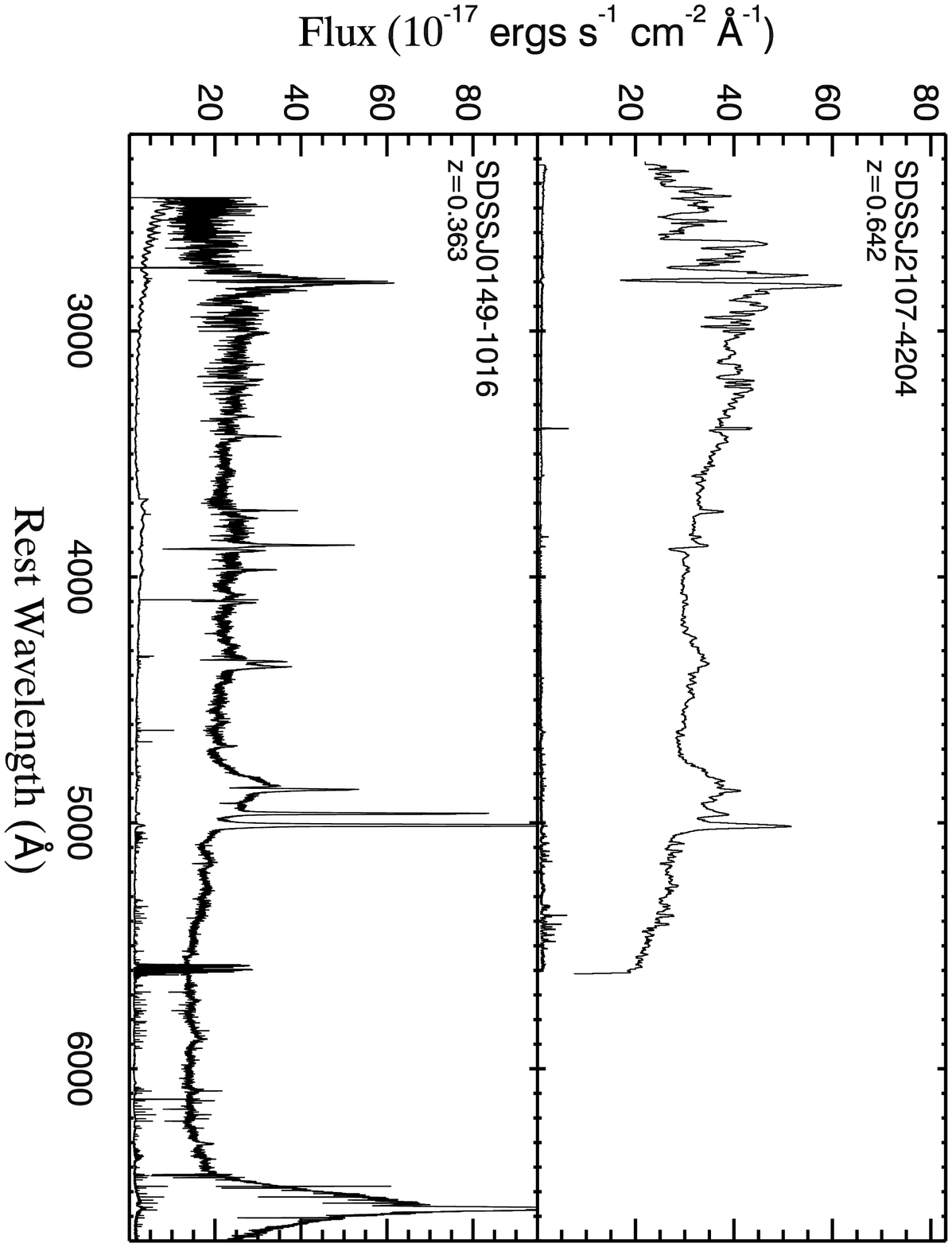}
\\ \footnotesize Fig.~2b. $-$ Cont.
\label{f2b}
\end{figure}

%

We also include in our study the object SDSS~J0149$-$1016 
(R.A. 01:49:06.736, decl. -10:16:49.27). While this 
object has a redshift of 0.364 and is therefore not included in our 
search statistics, we have a Magellan Inamori Kyocera Echelle (MIKE) 
spectrum of this quasar (also shown in Figure 2). MIKE is a 
high throughput double echelle spectrograph 
\citep[for details see][]{2003SPIE.4841.1694B}. The spectrum was 
taken on Sept. 15, 2006 and has been reduced with the standard 
MIKE Interactive Data Language (IDL) reduction tools. The continuum 
flux levels agree with the SDSS spectral observation (Sept. 22, 2001). 
SDSS~J0149$-$1016 was observed with a high resolution spectrograph with 
the intent of determining the outflow's radial distance. This work is 
still in progress. In the meantime, this spectrum provides an excellent
opportunity to compare the reddening determinations from the combinations
of H$\alpha$, H$\beta$, and H$\gamma$ NELs with the continuum as the MIKE 
spectrum has a larger spectral range ($\sim$3600$-$9300 \AA) than SDSS
that provides coverage of the Balmer series as well as the \mgii\ BAL 
and \feii\ absorption troughs (see Figure \ref{f1}). 

To strengthen the continuum reddening determinations in \S4, we 
complement the SDSS spectra with data from 
the 2 Micron All Sky Survey \citep[2MASS;][]{1997ASSL..210...25S}.
These IR data provide a longer baseline for the reddening
determinations. We use a matching radius of 3\arcsec~and obtain the 
J, H, and K magnitudes for each object. We list the 2MASS magnitudes 
for the sample in Table 1. The uncertainties for 2MASS data are 
typically 0.10$-$0.15 magnitudes.

\section{Reddening Determinations}

\subsection{Continuum Reddening}

We use the SDSS spectra to determine the continuum reddening rather 
than SDSS photometry because the combination 
of varying BEL emission line fluxes, and more importantly, the presence 
of BALs can significantly impact the E(B$-$V) determinations. We begin 
by correcting for Milky Way extinction (E(B$-$V)$_{MW}$) as determined 
by the online extinction calculator at the NASA/IPAC Extragalactic 
Database \citep{2011ApJ...737..103S}. Next, we correct for the intrinsic 
reddening of the object (E(B$-$V)$_{Q}$) in the restframe of the quasar 
using an SMC extinction curve (see Section 1; Richard et al. 2003; 
Hopkins et al. 2004). Because the SMC curve may not provide the most 
accurate description of the reddening at short wavelengths (i.e., in 
the far UV regime; see Dunn et al. 2010), we fit the restframe spectrum 
in the near UV and visible regimes, which are less dependent on the 
dust grain size distribution. To represent a typical unreddened quasar 
spectrum, we use the SDSS composite spectrum \citep{2001AJ....122..549V}. 
With regards to the intrinsic shape of the full spectral energy 
distribution (SED), Vanden Berk et al. (2001) and Richards et al. (2003) 
found that the distribution of quasar SED shapes from SDSS data is 
Gaussian with a relatively small spread ($\sigma$ = 0.30). In addition, 
\citet{2004AJ....128.1112H} showed that large reddening (E(B$-$V) $>$ 0.1) 
occured in less than 1\% of their sample of SDSS objects, in agreement 
with other surveys such as \citet{2009ApJ...698.1095U} and 
\citet{2012ApJ...757...51G}.Thus, any large deviation from the composite 
SDSS spectrum is likely due to dust extinction. 

\begin{figure}[!h]
  \centering \includegraphics[angle=90,width=0.9\textwidth]
  {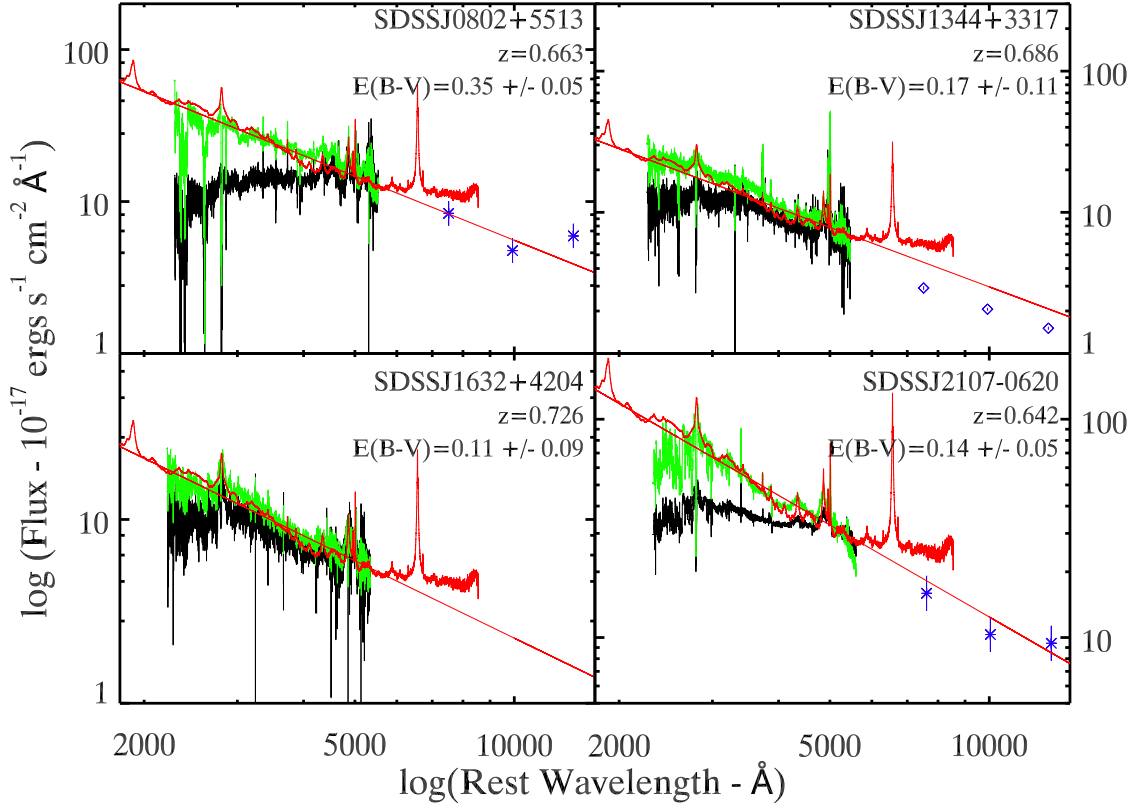}\\
  \caption[]
  {\footnotesize Spectral correction for both Milky Way and host
galaxy reddening for the FeLoBAL objects in the SDSS survey. The 
black histogram is the SDSS spectrum, the red histogram is the 
SDSS composite spectrum, the red curve is the ``UV-soft'' SED 
adjusted to the SDSS flux at 5100 \AA, the green histogram is 
the dereddened spectrum, and the blue asterisks and diamonds 
are the flux values and uncertainties derived from the 2MASS and 
UKIDSS photometry, respectively. The SDSS composite spectrum 
likely deviates from the SED (and the 2MASS data) longward of 
5000 \AA\ due to host galaxy contamination in the composite 
spectrum \citep{2001AJ....122..549V}. The derived values of 
E(B$-$V) are listed along with the redshift of each object in 
the upper right. Note that the data near 4500 \AA\ for 
SDSS~J0802+5513 are poor matches to the SDSS spectrum, 
primarily due to strong \feii\ emission (similar to EV1 of 
\citealt{1992ApJS...80..109B}). The poor fit to the UKIDSS data
for SDSS~J1344+3317 is plausibly due to a 10\% decrease in flux 
of the object in the time span between the observations (approx. 
2 years) given the similar slopes.}
  \label{f3}
\end{figure}

We de-redden each LoBAL spectrum in the sample by the SMC curve to match 
the continuum levels of the SDSS composite spectrum. We complement the SDSS 
spectrum with 2MASS photometry, when available, which provides data over 
a larger wavelength baseline that are less affected by reddening for the 
fit. The best fit is determined by matching regions of continuum between 
the two spectra shortward of approximately 5000 \aa. Due to the differences 
in the intrinsic continuum slope, presence of emission lines, and various 
depths and velocities of the BALs, we determine fits via visual inspection. 
While these fits are subject to the possiblity of host galaxy contamination 
in the quasars' spectra, a lack of strong stellar absorption troughs  
in the restframes of the quasars suggests that the host galaxy 
contributions are relatively small. 
We derive limits on uncertainty from under and over compensating 
for the reddening given the limits of the spectral noise in the LoBAL's 
spectrum. We also model the general shape of the continuum to provide 
predictions of flux levels in the infrafed range with both a 
\citet{1987ApJ...323..456M} SED and the ``UV-soft'' SED of Dunn et al. 
(2010) and find little difference in the fits between the two 
SEDs across the observed range. We show examples of the FeLoBAL fits in 
Figure \ref{f3} and 
list the E(B$-$V)$_{MW}$ values, determined quasar extinction values 
(E(B$-$V)$_Q$), and uncertainties for the LoBALs in Table 2. Regarding 
SDSS~J0149$-$1016, because we have the Magellan MIKE and SDSS
spectra we measure the continuum reddening for both. We list
the SDSS measured value (0.10$\pm$0.02) in Table 2 and find a consistent
value from the Magellan spectrum of 0.11$\pm$0.03.

As stated in Section 2, the AGN SED slopes likely have a Gaussian 
distribution centered around the slope of the average spectrum. This
implies, especially in a sample of non-BAL objects, that a fraction 
of the objects will have bluer slopes than the SDSS composite spectrum. 
We observe this in a small, randomly selected subset of non-BAL quasars 
using our fitting technique as a negative E(B$-$V)$_{Q}$, which suggests 
that these objects have little or no reddening. Similar to the non-BAL 
quasars, we only find a few objects in the LoBAL sample with appreciable 
negative E(B$-$V)$_{Q}$ values. We summarize the relative percentages of 
LoBALs in several reddening ranges in Table 3. Approximately 49\% of the 
LoBALs 
in the sample have an E(B$-$V)$_{Q}$ consistent with zero and 23\% of 
the objects 
have signifcant reddening (E(B$-$V)$_{Q}\ge$0.1). This is a substantially 
larger fraction compared to a general sample of non-BAL quasars (1\% in the 
same range; Hopkins et al. 2004) and is consistent with the SDSS Data Release 
5 sample of \mgii\ LoBALs measured by Zhang et al. (2010). 


\subsection{NLR Reddening}

\begin{figure}[!h]
 \noindent
 \centering
 \begin{minipage}[t]{0.40\linewidth} 
  \vspace{-0.5cm}
  \includegraphics[angle=90,width=\textwidth]{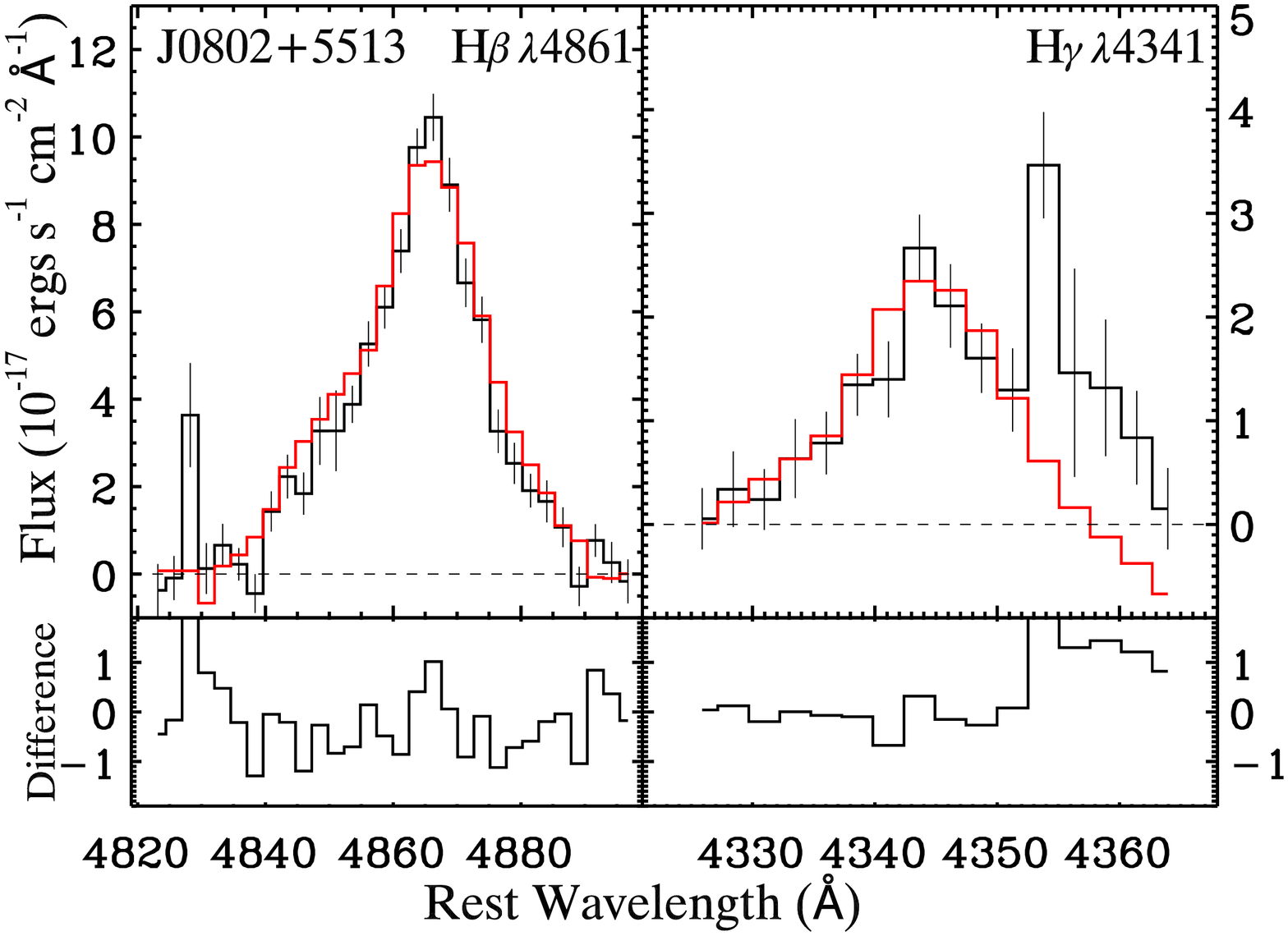}
 \end{minipage}
 \hspace{1.2cm}
 \begin{minipage}[t]{0.40\linewidth}
  \vspace{-0.5cm}
  \includegraphics[angle=90,width=\textwidth]{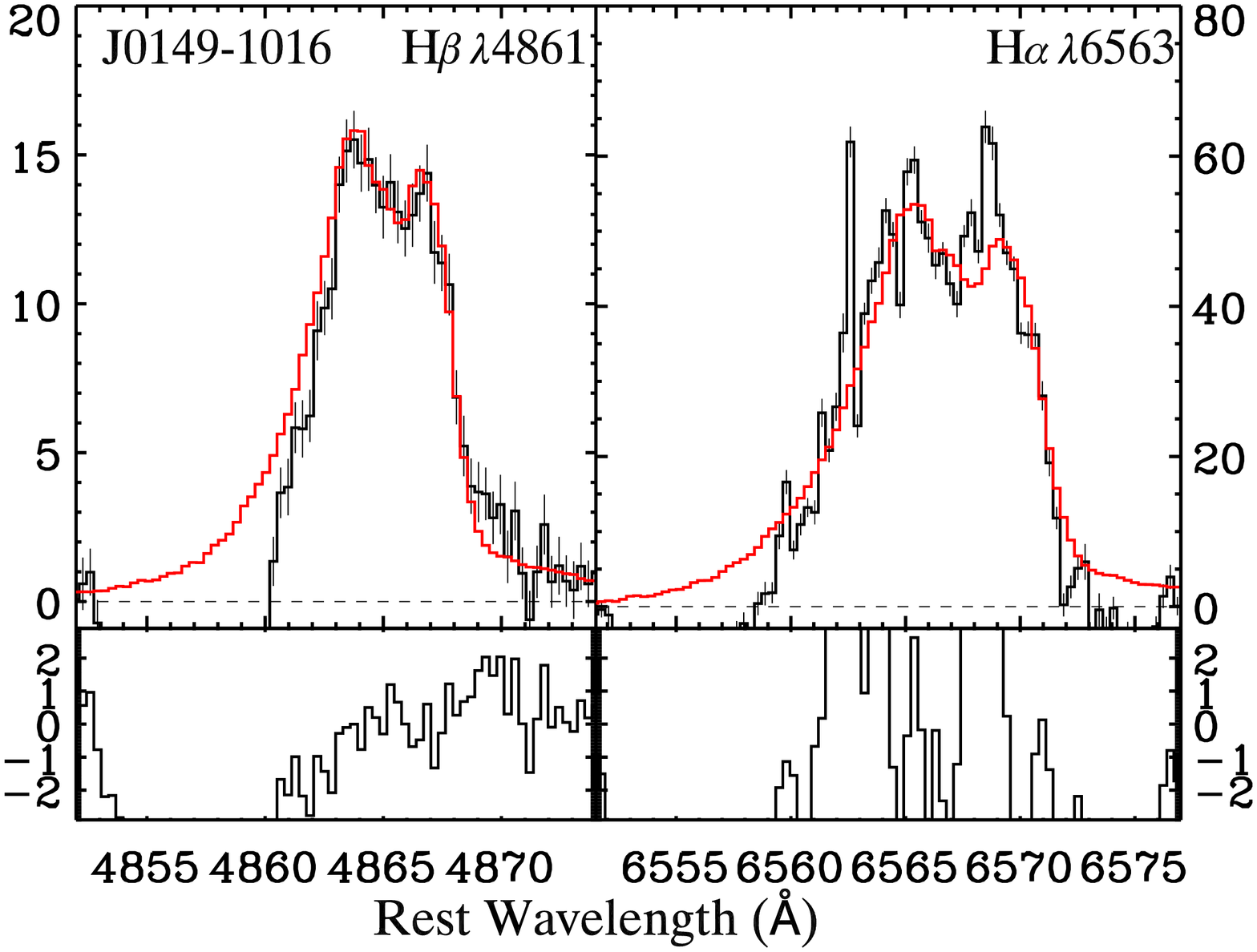}
\end{minipage}
\caption{\scriptsize {\bf Left:} Plots of the H$\beta$ $\lambda$4861 
and H$\gamma$ $\lambda$4341 narrow emission lines with their fits for
SDSS~J0802+5513. The black histograms are the hydrogen emission (after 
subtracting the
broad emission lines) and the red histograms are templates
created from the [\oiii] emission line ($\lambda$5007) scaled to
fit the hydrogen lines. The [\oiii] lines are boxcar smoothed by
5 pixels, which effectively removes problematic pixels due to
telluric line subtraction and maintains the general shape of
the [\oiii] line (FWHM = (800 $\pm$ 100) km/s). The poor fit to
the H$\gamma$ red wing is presumably due to contamination by
$[$\oiii$]$ $\lambda$4363. Thus, we fit the blue wing and peak.
Residual differences are ploted in the lower sections.
We determine uncertainties in fitting by taking the difference
of matching the peaks to matching the emission line wings, which
are comparable to the average of the residual differences (better
than 20\% across several pixels). {\bf Right:} Plot of the H$\alpha$
and H$\beta$ emission lines in SDSS~J0149$-$1016. The data is rebinned
by a factor of two and yields a ratio of 3.23 between H$\alpha$ and
H$\beta$. The data points blueward of the emission peak reflect the
BALs seen in the Balmer series. The largest differences are due to
spurious points in the data, which are from imperfect telluric
corrections.
}
\label{f4}
\end{figure}

To determine the reddening of the hydrogen NELs, we first normalize the 
continuum and the BEL (as we do not require any physical information from 
the BLR, see \S2) with a spline fit and subtract the fit to isolate the 
NEL. Using the [\oiii] $\lambda$5007 emission line as a template for the
NEL, we scale the template to match the normalized hydrogen NELs (as
demonstrated in Figure \ref{f4} for SDSS~J0802+5513). We use this template
primarily to determine the portion of the total emission profile that is
due to the NEL and to also account for any asymmetries in the line 
profile commonly found in NELs such as multiple peaks (e.g., 
SDSS~J1344+3317) and asymmetric blue wings
\citep{1981ApJ...247..403H,2005ApJ...630..122G,2008ApJ...680..926K}. We 
do not model the lines with Gaussian fits, which would require several 
fits that do not yield any physical information.
To determine the measurement uncertainties, we add the average flux 
uncertainty percentage across the peak of the line with the continuum 
determination uncertainty (approximately 3\%). These are listed in 
Table 2.

\noindent

For SDSS~J0149$-$1016 (also shown in Figure \ref{f4}), there are three 
important features. First, we fit H$\alpha$, H$\beta$, and H$\gamma$ 
due to the broader spectral coverage.
Second, this object also shows structure in the narrow emission lines, 
seen in both the [\oiii] doublet $\lambda\lambda$4959, 5007 and Balmer 
series. Finally, this object shows blueshifted Balmer absorption troughs, 
which implies a high density for the outflowing gas (see \S5 for further 
discussion of the physical implications). In the measurements of the 
emission features, the absorption troughs are important as they directly 
impact the blue emission wing for fitting. Therefore, our matches are 
primarily compared to the red wings and peaks of the emission lines.

In the case of SDSS~J1632+4204, we are unable to generate an adequate
template for the NEL. The [\oiii] doublet ($\lambda\lambda$4959, 5007) 
lies in the telluric absorption corrected region of the spectrum, which
provides significant errors in an [\oiii] template as well as any
measurement of the H$\beta$ emission. There is also no indication 
of a strong [\oii] $\lambda$3727 line to provide a template. Due to 
the absence of an uncontaminated line for the template but clear 
presence of H$\beta$ emission, we include this object in the survey 
statistics, but cannot fit the H$\beta$ line. 

\begin{figure}[!h]
 \noindent
 \centering
 \begin{minipage}[t]{0.40\linewidth} 
  \vspace{-0.5cm}
  \includegraphics[angle=90,width=\textwidth]{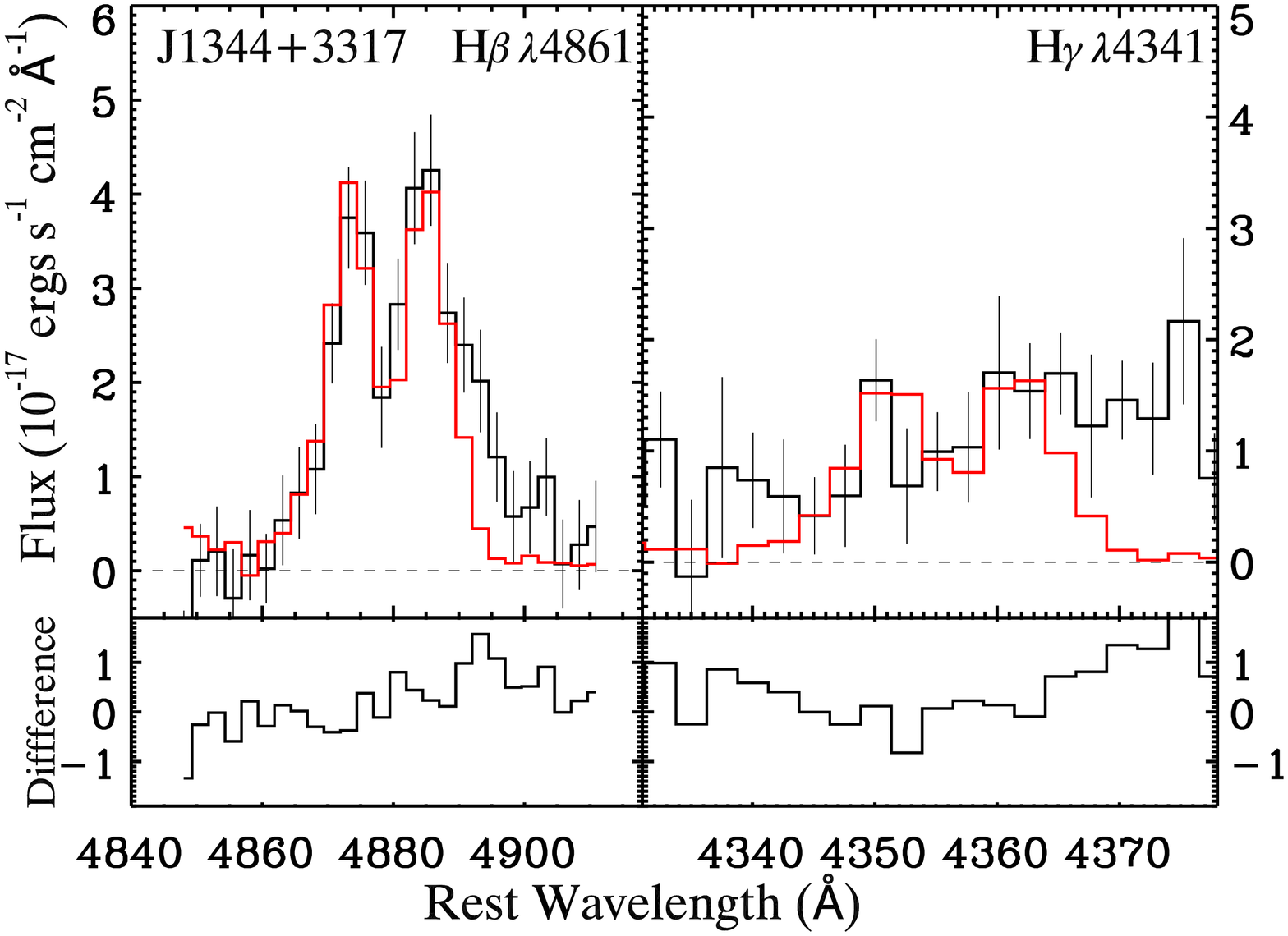}
 \end{minipage}
 \hspace{1.2cm}
 \begin{minipage}[t]{0.40\linewidth}
  \vspace{-0.5cm}
  \includegraphics[angle=90,width=\textwidth]{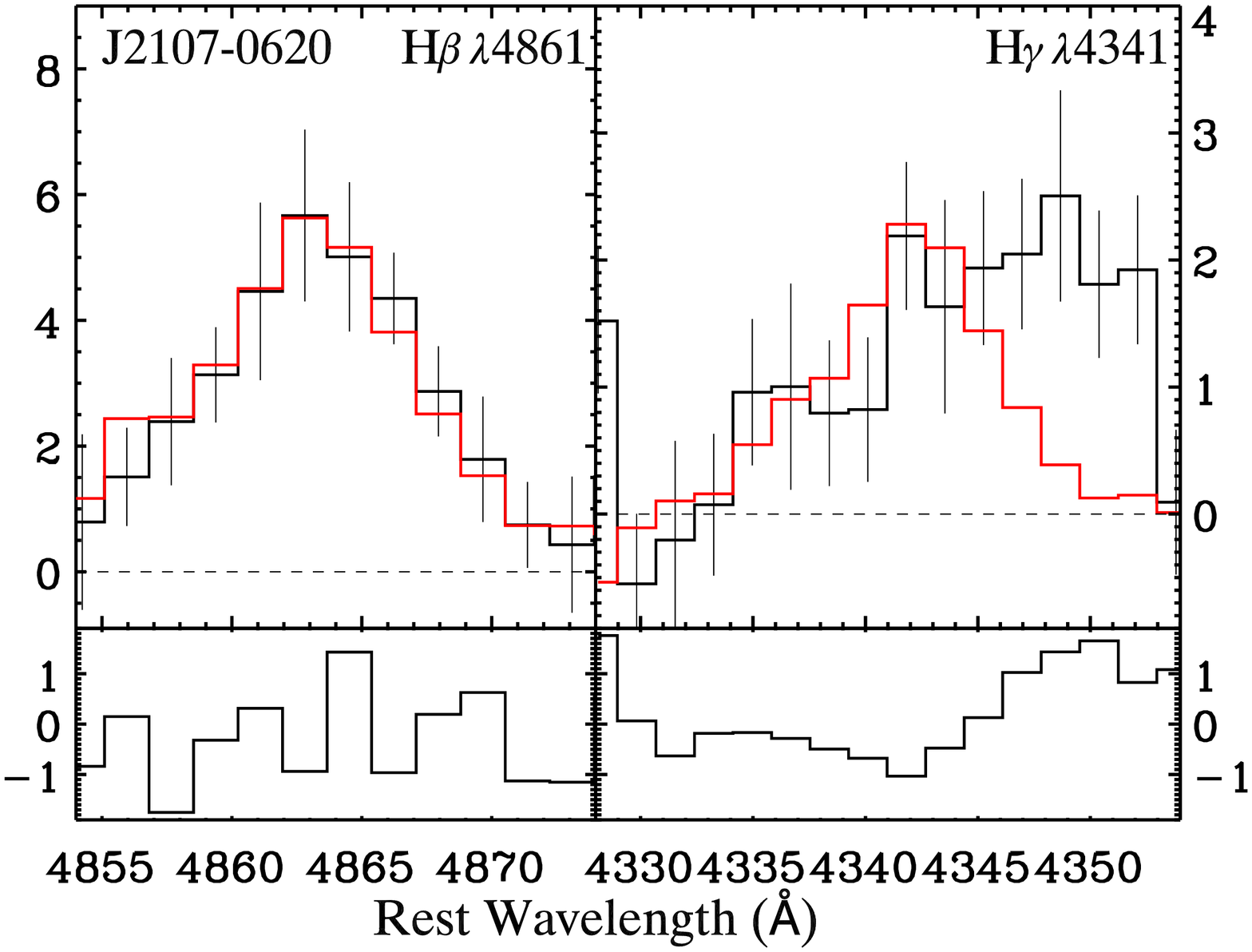}
\end{minipage}
\caption{\scriptsize {\bf Left:} Similar plot to Figure \ref{f4}, 
however for SDSS~J1344+3317. The fit to the H$\gamma$ emission, is a 
firm upper limit determined by the flux uncertainties. As with Figure 
\ref{f4}, the excess emission near 4370 \AA\ is likely due to 
$[$\oiii$]$ $\lambda$4363. No boxcar smoothing was applied to the 
[\oiii] template for this object, as the emission lines in this object 
are free of telluric contamination. {\bf Right:} Plot for SDSS~J2107$-$0620.
Unlike with SDSS~J1344+3317, we boxcar smooth the oxygen line profile by
a factor of 3 as the line is much noisier in this object's spectrum. Also
peculiar in this spectrum, the oxygen line appears to be shifted by
approximately 2 \AA\ from the Balmer lines, but does not affect the
overall determined ratio.}
\label{f5}
\end{figure}

As previously identified in Section 3, the only object in the SDSS survey
that exhibits a measurable narrow emission line from H$\gamma$ is
SDSS~J0802+5513. To determine the NLR reddening values in the remaining
two objects with measurable H$\beta$ emission lines, we use the template
to determine the upper limit of H$\gamma$ emission based on the statistical
error of the data (similar to the method used in Dunn et al. 2010 for
determining absorption trough limits). We compare a scaled emission line
template to the continuum region where the line would be detected. We 
maximize  the scale to match the amount of noise present in the spectral 
region, which provides the largest possible amount of emission present 
from H$\gamma$. We show this for SDSS~J1344+3317 and SDSS~J2107$-$0620 
in Figure \ref{f5}. Both objects show some signs of a weak and noisy
H$\gamma$ line, but we conservatively list these measurements as limits.
Finally, we list the E(B$-$V)$_{NLR}$ determinations and limits for these
two objects in Table 2 for each object in the sample.

\noindent

While the major focus of this study pertains to FeLoBALs due to the recent
distance determinations for the outflows of this type and the problems 
associated with reddening, we have also measured the NEL reddening for 
the regular LoBALs using the techniques listed above for the FeLoBALs. We 
list the values of E(B$-$V)$_{NLR}$ for the LoBALs in Table 2. 

\subsection{Reddening Comparisons}

Many studies of NEL/BEL/continuum reddening in AGN (e.g., Reynolds et
al. 1997) compare hydrogen column densities ($N_H$) between components 
of the unified model. However, \citet{2001A&A...365...37M}
showed that the E(B$-$V)/$N_H$ ratio ranges by two orders of
magnitude depending on the type of dust in the object. It has also 
been shown (Dunn et al. 2010; Hall et al. 2004) that while the 
reddening curves in many AGN are best fit by an 
SMC curve, in extreme cases of reddening the continuum is not well 
fit at short wavelengths. This disparity potentially affects any dust
column density determinations. Thus, we directly compare E(B$-$V) 
values between the AGN components.

Comparing the values for the FeLoBALs in Table 2, 
we find that three objects show clear evidence of reddening in both 
the continuum and the NELs (SDSS~J0802+5513, SDSS~J1344+3317, and 
SDSS~J2107$-$0620). While SDSS~J0149$-$1016 shows measurable
continuum reddening, due to the uncertainties in the line fitting,
the ratios and the resulting E(B$-$V)$_{NEL}$ values are consistent 
with either reddening scenario. As stated previously, the NELs 
in SDSS~J1632+4204 cannot be measured, but the object does show an 
appreciable continuum reddening.

All three FeLoBALs with measurable H$\beta$ NELs listed above,  
demonstrate a dusty source that simultaneously reddens both the NLR and 
the continuum. This implies that the dust must exist radially 
exterior to the NLR and precludes a torus reddening source for these 
three FeLoBALs. Due to the relatively low number of available objects 
and the combination of low continuum reddening (E(B$-$V)$_Q\approx$0.1) 
with large uncertainties in SDSS~J0149$-$1016, we cannot state 
conclusively if FeLoBALs in general share this physical picture. It 
is notable, though, that all three objects with significant continuum 
reddening appear to have reddening sources farther from the central
black hole than the NLR.

For the regular LoBALs, the majority of the LoBALs have E(B$-$V)$_{Q}$ 
values consistent with no reddening; we find that the majority of 
the E(B$-$V)$_{NLR}$ values are likewise small. There are 6 LoBALs 
of the 70 in the survey with E(B$-$V)$_{Q}$ $>$ 0.1. Of these 6 
objects, three have measurable H$\beta$ lines: SDSS~J1010+1843, 
SDSS~J1700+3955, and SDSS~J1703+3839. We find an 
E(B$-$V)$_{Q}$=0.11$\pm$0.03 for SDSS~J1010+1843. The emission line 
measurements yield a similar value though with a significant uncertainty. 
SDSS~J1700+3955 has moderate continuum reddening with a relatively large 
uncertainty (E(B$-$V)$_{Q}$=0.16$\pm$0.14). The measured NLR reddening 
is consistent with the continuum within the large
uncertainties of the two values (E(B$-$V)$_{Q}$=0.59$\pm$0.30). 
Finally, SDSS~J1703+3839 has a large reddening value of 
E(B$-$V)$_{Q}$=0.40$\pm$0.03, but unfortunately has no measurable 
H$\gamma$ emission line. The limit derived from H$\gamma$ yields an 
E(B$-$V)$_{NLR}>$0.2 and is certainly in line with the continuum 
determination. Simlar to the FeLoBALs, the LoBALs in the survey also have 
continuum and NLR reddenings that are similar within the uncertainties, 
which tentatively suggests that the reddening sources in LoBALs exist at 
larger distances than their NLRs. 

\begin{figure}[!h]
  \centering \includegraphics[angle=90,width=0.8\textwidth]
  {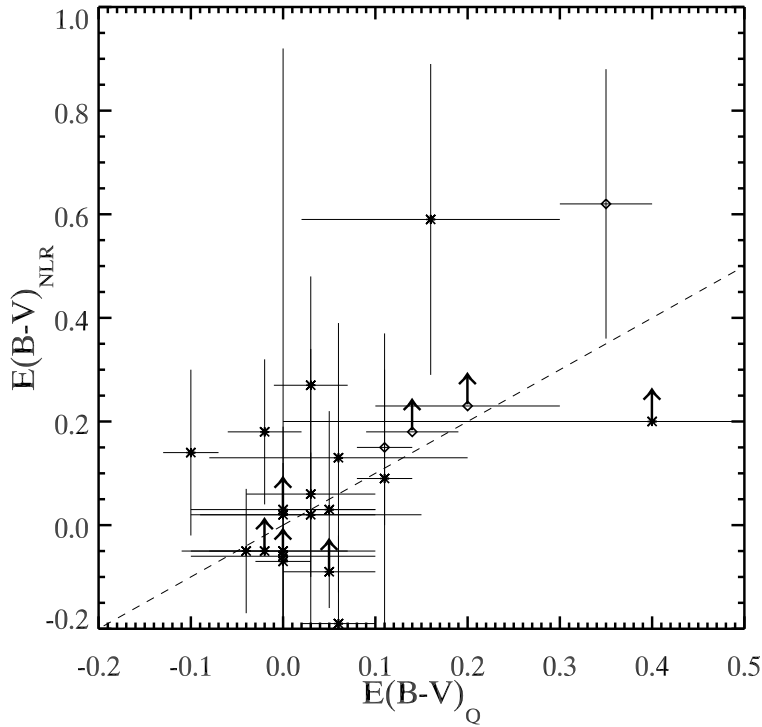}\\
  \caption[]
  {\footnotesize Comparison of the reddening for the quasar's
continuum (horizontal axis) and the reddening of the NLR (vertical
axis). Diamonds represent the FeLoBALs, while x's represent 
LoBALs. Vertical arrows illustrate objects with a lower limit 
determination for the NLR reddening.}
  \label{f6}
\end{figure}

To illustrate the relationship, we plot in Figure \ref{f6} the 
continuum reddening against the NLR reddening for both the LoBAL and 
FeLoBAL quasars in the entire sample. The plot shows a significant 
agreement between the two reddenings determinations for both populations 
of quasars. This nearly 1-to-1 relationship supports scenario 2 for the 
physical picture for the quasars in this sample.

\section{Summary and Conclusions}

We have assembled a subsample of 70 LoBAL quasars from a sample of
8,511 quasars with redshifts between 0.45$\ge$z$\ge$0.80 from the SDSS.
Of the 70 quasars, 23 have detectable narrow emission lines from
H$\beta$. This sample also contains 15 FeLoBAL quasars, 4 of which show 
signs of narrow Balmer emission lines. Therefore, narrow Balmer emission 
only appears to arise in approximately 27\% of FeLoBAL quasars. Both the 
LoBAL sample and FeLoBAL subsample clearly have a large fraction of 
objects with significant continuum reddening ($\sim$23\% with 
E(B$-$V)$_{Q}\ge0.1$ for \mgii\ BALs and 4 \feii\ objects or $\sim$23\% 
with E(B$-$V)$_{Q}\ge0.1$).

From measurements of the reddening of the continuum and narrow line 
region, we find that three FeLoBAL quasars (out of 5 
total objects, 4 from our sample and one additional quasar with a 
Magellan MIKE spectrum) have reddening sources that must exist radially 
exterior to the NLR from the AGN and are not directly tied to the dust 
torus. Due to telluric contamination, the H$\beta$ Balmer emission in one 
object (SDSS~J1632+4204) cannot be measured. Although the final object 
shows NLR reddening that is consistent with a radially exterior source, 
due 
to relatively low reddening (E(B$-$V)$_{Q}\approx$0.1) and significant 
uncertainties in the emission line fits, we cannot conclusively state 
which physical picture applies. Also, as strong NELs do not 
appear to be ubiquitous to FeLoBALs, these objects could potentially
be inherently different from the general population.

Given that every measurable object in the FeLoBAL sample has a similar
physical picture, the findings hint that the reddening frequently
observed in FeLoBALs is due to a source situated outside of the outflow
from the AGN. In studies of FeLoBALs, (e.g., Moe et al. 2009; 
Dunn et al. 2010; Borguet et al. 2013) the outflow distance 
determinations 
have all yielded distances of several kpc. Because the objects in our 
FeLoBAL survey all suggest that the dust is located radially exterior 
to the outflow, the dust would likewise exist at kpc scale distances 
(assuming these objects are similar to previously studied FeLoBALs). 
A simple explanation for a host of the dust would be the outflows 
themselves. This could be further evidence that FeLoBAL quasars
are evolutionary states of disrupted systems associated with major 
galaxy mergers as suggested by Glikman et al. (2012) and others.

In the case of SDSS~J0149$-$1016, should the dust lie 
at a galactic scale, the implied outflow distance is likely 
significantly smaller than other measured FeLoBALs. This distance
is implied by the presence of hydrogen Balmer line troughs, which 
are uncommon in BAL quasars. Hydrogen Balmer absorption typically 
suggests a rather large number density (albeit this depends 
on the ionization state of the gas). Thus, the dust would not arise 
due to the outflow but due to a source beyond the NLR such as a 
galactic dust lane or a nearby dwarf galaxy not associated with the 
outflow itself in at least this one object.

We have also examined the continuum and NLR reddening values in the 
larger sample of LoBAL quasars. While many of these objects have low 
continuum reddening values (E(B$-$V)$_{Q} < $0.1), three objects exhibit 
relatively large reddening values. The two largest reddened objects both
have Balmer emission lines that appear to have similarly large reddening;
this suggests that like the FeLoBALs the dust exists at larger
distances than the NLR for LoBAL quasars.

Due to the rare nature of LoBAL and FeLoBAL quasars and the subsequent 
small sample size (especially with regards to the FeLoBAL quasars), the 
next step is to increase the number of targets. Obtaining spectra for 
higher redshift FeLoBALs observed by SDSS with higher resolution 
and larger wavelength coverage (similar to SDSS0149$-$1016) will
increase the sample size and test these findings. 
Infrared spectra will provide coverage of the restframe optical 
regime for the stronger H$\alpha$ and H$\beta$ emission lines and
yield a greater number of FeLoBALs with narrow Balmer emission 
suitable for the technique outlined in this study.

LCH acknowledges additional support from the Kavli Foundation, 
Peking University, the Chinese Academy of Sciences, and 
the Carnegie Institution for Science.

\input{table1}
\input{table2}
\input{table3}

%% file: table1.tex
\begin{deluxetable}{lcccccccc}
\rotate
\tablecolumns{9}
\tablewidth{1.2\textwidth}
\tabletypesize{\normalsize}
\tablecaption{SDSS Quasars with \mgii\ Outflows and Properties}
\tablehead{
\colhead{SDSS Name} &
\colhead{Plate-MJD-Fiber$^a$} &
\colhead{r' Mag} &
\colhead{J Mag$^c$} &
\colhead{H Mag$^c$} &
\colhead{K Mag$^c$} &
\colhead{z$^b$} &
\colhead{\feii$^d$} &
\colhead{H$\beta$$^e$} \\ 
} 
\startdata
J010352.46+003739.7 	& 51816-0396-471 & 17.42 & 16.157 & 15.425 & 14.074 &	0.703	&	& \\
J024220.10$-$085332.7 	& 51910-0456-291 & 18.86 &  - 	  &  - 	   &  -     &	0.799	&	& y     \\
J080248.19+551328.8 	& 53384-1871-440 & 18.42 & 16.434 & 15.923 & 14.645 &	0.663	& y 	& y	\\
J080611.02+115029.0 	& 53794-2418-616 & 18.88 &  - 	  &  - 	   &  -     &	0.616	& y 	& \\
J080934.64+254837.9 	& 52670-1205-588 & 17.93 & 17.029 & 16.469 & 16.099 &	0.545	&	& y     \\
J082231.53+231152.0 	& 53317-1926-546 & 17.68 &  - 	  &  - 	   &  -     &	0.653	&	& y     \\
J083525.98+435211.3 	& 52232-0762-175 & 17.55 & 16.407 & 15.961 & 15.131 &	0.568	&	& \\
J085053.12+445122.4 	& 52605-0897-359 & 17.41 & 16.222 & 15.967 & 14.909 &	0.542	&	& \\
J085357.88+463350.6 	& 52238-0764-248 & 18.22 & 16.758 & 15.892 & 15.287 &	0.549	&	& \\
J092525.16+202139.0 	& 53708-2289-183 & 18.51 & 17.067 & 16.080 & 15.288 &	0.460	&	& \\
J093228.56+555344.8 	& 51991-0556-251 & 18.76 & 16.611 & 15.820 & 14.826 &	0.876	&	& \\
J093315.09+351944.2 	& 52992-1594-348 & 18.87 &  - 	  &  - 	   &  -     &	0.525	&	& y     \\
J094225.42+565613.0 	& 52253-0557-359 & 18.52 &  - 	  &  - 	   &  -     &	0.831	&	& \\
J094443.13+062507.4 	& 52710-0993-535 & 16.24 & 14.831 & 14.517 & 13.808 &	0.695	&	& \\
J101038.76+184321.8 	& 53768-2373-124 & 18.97 & 16.998 & 16.123 & 15.433 &	0.745	&	& y     \\
J101729.52+264146.7 	& 53765-2350-435 & 18.70 &  - 	  &  - 	   &  -     &	0.526	&	& y     \\
J102802.33+592906.7 	& 52316-0559-582 & 18.92 &  - 	  &  - 	   &  -     &	0.535	&	& \\
J103036.93+312028.8 	& 53440-1959-121 & 17.69 & 16.368 & 15.815 & 15.059 &	0.873	&	& \\
J103255.37+083503.2 	& 52734-1240-316 & 17.73 & 15.750 & 15.641 & 15.229 &	0.891	& y 	& \\
J104122.84$-$005618.4 	& 51913-0274-060 & 18.50 & 16.951 & 16.176 & 15.358 &	0.497	&	& y     \\
J104210.43+501609.1 	& 52354-0875-583 & 18.07 & 16.650 & 16.238 & 15.600 &	0.787	&	& \\
J104459.60+365605.1 	& 53463-2090-329 & 16.81 & 15.537 & 15.157 & 14.331 &	0.701	& y 	& \\
J105259.99+065358.0 	& 52670-1001-080 & 18.56 &  - 	  & - 	   &  -     &	0.722	&	& \\
J105856.75+480805.4 	& 52646-0964-305 & 18.00 & 16.539 & 15.965 & 15.313 &	0.591	&	& y \\
J111628.00+434505.8 	& 53061-1364-095 & 17.44 & 15.887 & 15.748 & 15.291 &	0.801	&	& \\
J112526.12+002901.3 	& 51614-0281-427 & 18.10 & 16.485 & 16.300 & 15.356 &	0.864	& y 	& \\
J112621.25+343628.9 	& 53713-2100-539 & 17.86 & 16.545 & 16.308 & 15.594 &	0.603	&	& \\
J112822.42+482310.0 	& 52642-0966-203 & 17.59 & 16.546 & 16.265 & 15.919 &	0.543	&	& \\
J112828.31+011337.9 	& 51992-0512-123 & 18.60 & 16.055 & 15.386 & 14.440 &	0.893	& y 	& \\
J113807.83+531231.6 	& 52367-0880-404 & 18.67 & 16.953 & 16.406 & 16.107 &	0.790	&	& \\
J114043.62+532438.9 	& 52734-1015-085 & 18.31 & 16.942 & 16.408 & 15.304 &	0.530	&	& \\
J114209.01+070957.7 	& 53383-1621-306 & 18.42 &  - 	  &  - 	   &  -     &	0.497	&	& \\
J120751.48+253953.7 	& 54484-2656-420 & 18.18 & 16.579 & 16.361 & 15.802 &	0.874	&	& y     \\
J121113.38+121937.3 	& 53149-1612-179 & 18.60 & 16.764 & 15.894 & 14.893 &	0.464	&	& y     \\
J121303.40$-$014450.9 	& 52367-0332-579 & 18.54 &  - 	  &  - 	   &  -     &	0.612	&	& y     \\
J121442.30+280329.1 	& 53823-2229-557 & 17.38 & 15.790 & 15.649 & 14.594 &	0.695	& y 	&\\
J123820.19+175039.1 	& 54234-2599-503 & 16.42 & 15.326 & 14.285 & 12.957 &	0.453	&	& y     \\
J124300.87+153510.6 	& 53502-1769-584 & 18.52 &  - 	  &  - 	   &  -     &	0.561	& y 	&\\
J130741.12+503106.4 	& 52753-1281-361 & 18.25 &  - 	  &  - 	   &  -     &	0.701	&	&\\
J131433.19+471457.7 	& 53062-1461-431 & 18.88 & 15.518 & 15.017 & 14.606 &	0.869	&	&\\
J134415.75+331719.1 	& 53503-2024-346 & 18.82 &  - 	  &  - 	   &  -     &	0.686	& y 	& y	\\
J134651.31+302421.7 	& 53851-2094-598 & 18.44 &  - 	  &  - 	   &  -     &	0.867	&	&\\
J140806.20+305448.4 	& 53795-2125-236 & 17.41 & 16.090 & 15.646 & 14.993 &	0.830	&	&\\
J142649.24+032517.7 	& 52049-0584-004 &  8.37 & 16.723 & 16.015 & 15.176 &	0.530	&	&\\
J142927.28+523849.5 	& 52781-1327-343 & 17.51 & 16.257 & 15.585 & 14.845 &	0.594	&	&\\
J144211.79+533608.5 	& 52669-1163-293 & 18.67 &  - 	  &  - 	   &  -     &	0.863	& y 	&\\
J144800.15+404311.7 	& 53119-1397-198 & 17.09 & 15.118 & 14.903 & 14.322 &	0.801	& y 	&\\
J145233.68+250002.6 	& 54184-2143-603 & 18.50 &  - 	  &  - 	   &  -     &	0.587	&	&\\
J145724.00+452157.8 	& 53147-1676-449 & 18.62 & 17.094 & 16.074 & 15.447 &	0.717	&	&\\
J145736.70+523454.6 	& 52674-1164-184 & 18.08 &  - 	  &  - 	   &  -     &	0.637	&	&\\
J145836.73+433015.5 	& 52734-1290-601 & 18.59 & 16.926 & 16.630 & 15.491 &	0.761	&	&\\
J150847.41+340437.7 	& 53108-1385-173 & 17.35 & 15.690 & 15.696 & 14.859 &	0.798	&	& y     \\
J151306.52+200244.1 	& 54525-2156-258 & 18.85 &  - 	  &  - 	   &  -     &	0.703	& y 	&\\
J152350.42+391405.2 	& 52765-1293-234 & 16.65 & 15.347 & 14.866 & 13.860 &	0.661	&	&\\
J153209.51+061356.1 	& 54540-1819-024 & 17.06 & 15.545 & 15.251 & 14.892 &	0.835	&	& \\
J154351.92+162422.1 	& 54243-2518-350 & 16.24 & 14.672 & 14.301 & 13.224 &	0.849	&	& \\
J154620.98+453916.7 	& 52782-1333-601 & 17.57 & 16.274 & 16.134 & 15.594 &	0.459	&	& y     \\
J160143.75+150237.7 	& 54568-2524-356 & 17.26 & 15.772 & 15.638 & 14.587 &	0.650	& y 	& y	\\
J160234.88+160041.1 	& 53555-2197-225 & 18.18 &  - 	  &  - 	   &  -     &	0.719	&	& \\
J160656.95+133931.2 	& 54569-2527-346 & 18.67 &  - 	  &  - 	   &  -     &	0.452	&	& y     \\
J161637.15+390356.8 	& 52759-1336-088 & 17.90 & 16.970 & 16.333 & 15.344 &	0.810	&	&\\
J162435.28+090731.6 	& 54589-2532-378 & 18.06 & 16.547 & 15.936 & 14.949 &	0.652	& 	& y     \\
J163255.46+420407.8 	& 52379-0816-569 & 18.68 &  - 	  &  - 	   &  -     &	0.726	& y 	& y     \\
J163656.84+364340.4 	& 52782-1174-337 & 18.91 & 17.013 & 16.251 & 15.719 &	0.850	&	&\\
J164447.19+311437.2 	& 52781-1340-002 & 17.90 & 16.369 & 16.115 & 15.281 &	0.690	&	&\\
J170010.82+395545.8 	& 52079-0633-482 & 18.99 & 17.240 & 16.463 & 15.641 &	0.577	&	& y     \\
J170341.82+383944.7 	& 52071-0632-632 & 18.72 & 16.795 & 15.999 & 15.426 &	0.554	&	& y     \\
J204333.20$-$001104.2 	& 52435-0981-044 & 17.94 & 16.793 & 15.769 & 15.223 &	0.545	&	&\\
J210757.67$-$062010.6 	& 52174-0637-610 & 17.22 & 15.732 & 15.082 & 14.145 &	0.642	& y 	& y	\\
J220931.92+125814.5 	& 52519-0735-501 & 18.52 &  - 	  &  -     &  -     &	0.813	&       & y     \\
\enddata
\normalsize
 
\tablenotetext{a}{The corresponding SDSS plate number, modified julian date, and 
fiber number for the object.}

\tablenotetext{b}{  redshift of the object.}

\tablenotetext{c}{IR magnitudes from 2MASS.}

\tablenotetext{d}{Absorption troughs from \feii\ detected at similar velocities to \mgii.}

\tablenotetext{e}{BALs with narrow H$\beta$ emission lines.}
\end{deluxetable}

%% file: table2.tex
\newpage
\begin{deluxetable}{lccc}
\tablecolumns{4}
\tablewidth{0.6\textwidth}
\tablecaption{LoBAL Reddening Determinations}
\tablehead{
\colhead{SDSS Name} &
\colhead{E(B$-$V)$_{MW}$} &
\colhead{E(B$-$V)$_{Q}$}& 
\colhead{E(B$-$V)$_{NLR}$}\\ 
}
\startdata
FeLoBALs & & & \\
0149$-$1016$^a$ &0.038  &0.11$\pm$0.03 & 0.15$\pm$0.15 \\
0802+5513       &0.042  &0.35$\pm$0.05 & 0.62$\pm$0.26 \\
0806+1150       &0.022  &0.10$\pm$0.11  & $-$ \\
1032+0835       &0.021  &0.08$\pm$0.07  & $-$ \\
1044+3656       &0.013  &0.05$\pm$-0.04 & $-$ \\
1125+0029       &0.031  &0.11$\pm$0.09  & $-$ \\
1128+0113       &0.029  &0.10$\pm$0.10  & $-$ \\
1214+2803       &0.021  &0.08$\pm$0.07  & $-$ \\
1243+1535       &0.032  &0.00$\pm$0.10  & $-$ \\
1344+3317       &0.014  &0.20$\pm$0.10 & $>$0.23 \\
1442+5336       &0.009  &-0.05$\pm$0.05 & $-$ \\
1448+4043       &0.011  &-0.02$\pm$0.09 & $-$ \\
1513+2002       &0.032  &0.11$\pm$0.05  & $-$ \\
1601+1502       &0.042  &0.06$\pm$0.04  & $-$ \\
1632+4204       &0.025  &0.11$\pm$0.09 & $-^b$ \\
2107$-$0620     &0.058  &0.14$\pm$0.05 & $>$0.18 \\
 & & & \\
\hline
LoBALs & & & \\
0103+0037 	&0.029	&0.13$\pm$0.06  & $-$ \\
0242$-$0853	&0.022	&0.03$\pm$0.07  & 0.06$\pm$0.28 \\
0809+2548 	&0.032	&-0.02$\pm$0.09 & $> -$0.05 \\
0822+2311	&0.034	&0.00$\pm$0.03  & $> -$0.07 \\
0835+4352 	&0.028	&0.00$\pm$0.04  & $-$ \\
0850+4451 	&0.024	&-0.03$\pm$0.07 & $-$ \\
0853+4633 	&0.021	&0.00$\pm$0.03  & $-$ \\
0925+2021 	&0.038	&-0.05$\pm$0.15 & $-$ \\
0932+5553 	&0.025	&0.15$\pm$0.05  & $-$ \\
0933+3519 	&0.012	&0.00$\pm$0.10  & -0.06$\pm$0.18 \\
0942+5656 	&0.017	&0.06$\pm$0.04  & $-$ \\
0944+0625 	&0.026	&0.03$\pm$0.03  & $-$ \\
1010+1843 	&0.027	&0.11$\pm$0.03  & 0.09$\pm$0.28 \\
1017+2641 	&0.023	&0.00$\pm$0.10  & $>$ 0.03 \\
1028+5929 	&0.007	&0.00$\pm$0.10  & $-$ \\
1030+3120 	&0.018	&0.00$\pm$0.05  & $-$ \\
1041$-$0056 	&0.046	&-0.10$\pm$0.03 & 0.14$\pm$0.16\\
1042+5016 	&0.013	&0.03$\pm$0.05  & $-$ \\
1052+0653	&0.033	&0.00$\pm$0.09  & $-$ \\
1058+4808 	&0.011	&-0.02$\pm$0.04 & 0.18$\pm$0.14 \\
1116+4345 	&0.013	&0.06$\pm$0.04  & $-$ \\
1126+3436 	&0.022	&-0.03$\pm$0.03 & $-$ \\
1128+4823       &0.017  &0.05$\pm$0.04  & $-$ \\
1138+5312 	&0.007	&0.11$\pm$0.09  & $-$ \\
1140+5324 	&0.009	&-0.05$\pm$0.15 & $-$ \\
1142+0709 	&0.034	&0.00$\pm$0.10  & $-$ \\
1207+2539 	&0.018	&0.05$\pm$0.05  & 0.03$\pm$0.19 \\
1211+1219 	&0.025	&0.06$\pm$0.14  & 0.13$\pm$0.26 \\
1213$-$0144	&0.020	&0.03$\pm$0.04  & 0.27$\pm$0.21 \\
1238+1750 	&0.021	&-0.04$\pm$0.04 & $-$0.05$\pm$0.12 \\
1307+5031	&0.013	&0.08$\pm$0.07  & $-$ \\
1314+4714 	&0.009	&0.03$\pm$0.07  & $-$ \\
1346+3024 	&0.017	&0.00$\pm$0.03  & $-$ \\
1408+3054 	&0.009	&0.05$\pm$0.05  & $-$ \\
1426+0325 	&0.031	&-0.02$\pm$0.07 & $-$ \\
1429+5238 	&0.012	&0.02$\pm$0.08  & $-$ \\
1452+2500 	&0.030	&-0.03$\pm$0.13 & $-$ \\
1457+4521	&0.015	&0.02$\pm$0.05  & $-$ \\
1457+5234 	&0.016	&-0.10$\pm$0.11 & $-$ \\
1458+4330 	&0.014	&0.08$\pm$0.12  & $-$ \\
1508+3404 	&0.014	&0.06$\pm$0.04  & -0.19$\pm$0.17 \\
1523+3914 	&0.018	&0.04$\pm$0.03  & $-$ \\
1532+0613 	&0.042	&0.02$\pm$0.03  & $-$ \\
1543+1624 	&0.028	&0.06$\pm$0.04  & $-$ \\
1546+4539 	&0.013	&0.00$\pm$0.10  & $-$0.05$\pm$0.24 \\
1602+1600       &0.028  &0.02$\pm$0.05  & $-$ \\
1606+1339 	&0.036	&0.03$\pm$0.12  & 0.02$\pm$0.12 \\
1616+3903 	&0.007	&0.00$\pm$0.10  & $-$ \\
1624+0907	&0.057	&0.00$\pm$0.10  & 0.02$\pm$0.90 \\
1636+3643 	&0.012	&0.10$\pm$0.10  & $-$ \\
1644+3114 	&0.024	&0.23$\pm$0.07  & $-$ \\
1700+3955 	&0.020	&0.16$\pm$0.14  & 0.59$\pm$0.30$^d$ \\
1703+3839 	&0.036	&0.40$\pm$0.40  & $>$0.2 \\
2043$-$0011 	&0.056	&-0.02$\pm$0.08 & $-$ \\
2209+1258 	&0.072	&0.05$\pm$0.05  & $>$-0.09 \\
\enddata
\normalsize


\tablenotetext{a}{Object also has H$\alpha$. This ratio is between H$\beta$ to H$\alpha$.}

\tablenotetext{b}{Object has no uncontaminated \oiii\ emission template available.}

\tablenotetext{c}{Spectrum too noisy for acurate measurement.}

\tablenotetext{d}{Used H$\beta$ for line template.}

\end{deluxetable}

%% file: table3.tex
\newpage
\begin{deluxetable}{lc}
\tablecolumns{2}
\tablewidth{0.5\textwidth}
\tablecaption{Reddening Statistics}
\tablehead{
\colhead{E(B-V)$_{Q}$ Range} &
\colhead{LoBAL \%} \\
}
\startdata
E(B-V) $< -$0.10         & 0.0  \\
$-$0.10 $\ge$ E(B-V)$ < -$0.05 & 7.1  \\
$-$0.05 $<$ E(B-V) $<$ 0.05  & 48.6 \\
0.05 $<$ E(B-V) $<$ 0.09    & 25.7 \\
E(B-V) $\ge$ 0.10          & 22.9 \\
\enddata
\normalsize

\end{deluxetable}